\documentclass[5p,authoryear,times]{elsarticle}

\usepackage{graphicx}

\usepackage{amssymb}
\usepackage{lineno}
\usepackage{lscape}
\usepackage{tabularx}
\usepackage[gen]{eurosym}
\usepackage{multirow}
\usepackage{hhline}
\usepackage{booktabs}
\usepackage{siunitx}
\usepackage{framed}
\newcolumntype{L}{@{}l@{}} 

\journal{IEEE Software}

\begin{document}

\begin{frontmatter}

\title{The perceived effects of group developmental psychology training on agile software development teams}

\author[1]{Lucas Gren\corref{cor1}}
\ead{lucas.gren@cse.gu.se}
\cortext[cor1]{Corresponding author. Tel.: +46 739 882 010}

\author[2]{Alfredo Goldman}
\ead{gold@ime.usp.br}

\author[3]{Christian Jacobsson}
\ead{christian.jacobsson@psy.gu.se}

\address[1]{The Department of Computer Science and Engineering, Chalmers University of Technology and the University of Gothenburg, Gothenburg, Sweden}
\address[2]{The Department of Mathematics and Statistics, The University of S\~ao Paulo, S\~ao Paulo, Brazil}
\address[3]{The Department of Psychology, The University of Gothenburg, Gothenburg, Sweden}

\begin{abstract}
Research has shown that the maturity of small workgroups from a psychological perspective is intimately connected to team agility. We therefore tested if agile team members appreciated group development psychology training. Our results show that the participating teams seem to have a very positive view of group development training and state that they now have a new way of thinking about teamwork and new tools to deal with team-related problems. We therefore see huge potential in training agile teams in group development psychology since the positive effects might span over the entire software development organization.

\end{abstract}


\end{frontmatter}

\section{Introduction}\label{sec:introduction}
A few studies have been conducted that set out to investigate social-psychological aspects of agile development. \citet{whit}, for example, verify that agile teams need to look at social-psychological aspects to fully understand how they function. There are also a set of studies connecting agile methods to organizational culture (e.g.\ \citet{iivari}). These connect the agile adoption process to organizational culture and showed that there are cultural factors that could jeopardize the agile implementation. A more recent study has underlined the importance of focusing even more on social-psychological aspects of work groups (or teams) in software engineering in order to gain more descriptive and predictive power \citep{lenberg2015}.


\begin{framed}
\scriptsize \textbf{Sidebar 1: Basics on Groups and Teams} \citet{grupp} defines a group as three or more members that interact with each other to perform a number of tasks and achieve a set of common goals. This means that many large groups are in fact a set of smaller subgroups and should be handled as separate groups. \citet{wheelan2012} defines a team in an organization as a small workgroup that has common goals and effective methods to reach them. If the group consists of more than eight individuals, they are less productive than smaller groups \citep{wheelan2009}.
\end{framed}

\section{Related Work}
\subsection{Group development over time}
The study of the behavior of small groups was launched with the establishment of a research center of group dynamics in 1945, and several research groups proposed different ways of analyzing the behavior of groups \citep{wheelan2012}. Some studies propose group development can be described as states or levels of activity but an integrated theory of linear and cyclic models
was first introduced in 1965 by \citet{tuckman1965developmental}. The result of his analysis was a conceptual model including four stages of group development, namely, Forming, Storming, Norming, and Performing. The model suggested by \citet{wheelan2012} largely overlaps the stages from that model. A group at the later stages are described as being more ``mature,'' which is referred to as group (or team) maturity. 

\begin{figure}
\centerline{\includegraphics[width=90mm]{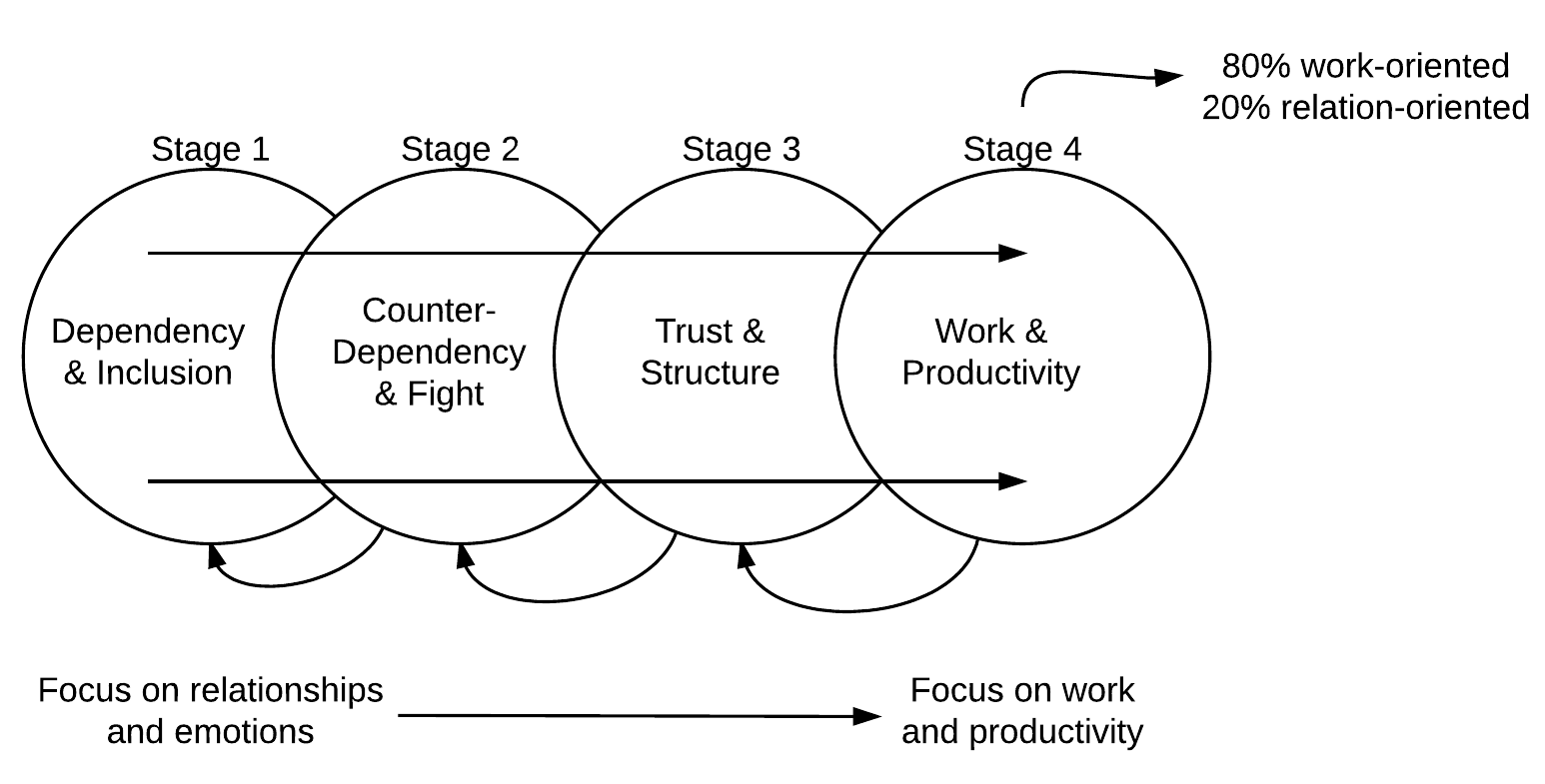}}
\caption{The Group Development Stages. Adopted from~\citet{wheelan2012}.}
\label{fig:groupstages}
\end{figure}

\begin{framed}
\scriptsize \textbf{Sidebar 2: The Integrated Model of Group Development.} In the Integrated Model of Group Development (or IMGD) groups develop across different maturity stages. This is straight-forward, since we all know we behave differently with people we do not know and people we do know. The developmental levels of groups can be compared to that of a human; we figure out what world we got born into (being a child), then we question the structures we see (adolescence), and finally we somewhat find our place in this world and can focus more on how to develop and mature \citep{wheelan2012}. The stages are shown in Figure~\ref{fig:groupstages} and are the following: 

\paragraph{Stage 1: Dependency and inclusion}
The first stage is categorized by three main areas; concerns about safety and inclusion, member dependency on the designated leader, and a wish for order and structure. The group is supposed to become organized, capable of efficient work, and achieve goals, so the first stage must have a purpose in getting there~\citep{wheelan2012}. 

\paragraph{Stage 2: Counter-dependency and fight}
When the group safely navigated through the previous stage, they have gained a sense of loyalty. As people feel more safety they will dare to speak up and express opinions that might not be shared by all members. The second stage of a group's development is therefore a conflict phase where fight is a must in order to create clear roles to be able to work together in a constructive way. The members have to go through this in order to be able to trust each other and the leader~\citep{wheelan2012}.

\paragraph{Stage 3: Trust and structure}
The third stage is a structure-developing phase where the roles are based on competence instead of striving for power or safety. Communication will be more open and task-oriented. The third stage of group development is characterized by more mature negotiations about roles, organization, and processes~\citep{wheelan2012}. 

\paragraph{Stage 4: Work and productivity}
The fourth and final stage (excluding the termination phase) is when the group wants to get the task done well at the same time as the group cohesion is maintained over a long period of time. The group also focuses on decision-making and encourages task-related conflicts. This is a time of intense productivity and effectiveness and it is at this stage the group becomes a team~\citep{wheelan2012}. 
\end{framed}

\subsection{Agility and group development}
There are many advantages of looking at group level instead of only individuals and their traits \citep{hogg2000we}. We can often find better explanations and more accurate models for our empirical data on other levels of abstraction \citep{hackman2003learning}, which is the case when treating the team as the unit of analysis. This is also verified by the few articles found within agile software development where more can be explained on a group (or team) level \citep{grenjss2} than on the individual level \citep{gren2018non}. The self-organizing, empowerment, and accountability properties of an agile team has been shown to be tightly linked to the higher group maturity stages \citep{grenjss2}. Agile software development is now a very common approach to projects that has been shown to increase project success \citep{serrador2015does}. In a study by \citet{mcdonald}, also in the software engineering domain, the authors conclude that as much is derived from the team's context, as of the people in it. The term \emph{team} is almost used exclusively for small workgroups in software engineering research and practice. However, in social and organizational psychology, a distinction is sometime made between work-group and teams and the difference is that a team is a well functioning work-group that has reached a more mature stage of their group development. The personality research in software engineering has not provided coherent results \citep{forty}, which also indicated that further studies should look at the team-level as the unit of analysis instead. Furthermore, studies have shown that personality tests have little predictive value (see e.g. \citet{licorish2015communication}).

Since group maturity has been shown to be important and connected to team agility \citep{grenjss2}, we conducted group developmental psychology training for agile teams and handed out surveys to ask them of the usefulness of thinking about the team as an entity that develops over time and knowing the general patterns that all human groups must go through. For an easy read and an introduction to the group development model, see \citet{wheelan2012}. Our research question was: ``Are there any perceived effects of a 1.5-hour group development psychology training on agile software development teams?''

\section{Method}
We met 12 agile teams from two companies in Brazil, and the group sizes ranged from 3 to 11 members for the participating groups and comprised 85 group members. The teams from the companies were from the IT departments at one large on-line media and social networking enterprise with around 5,000 employees, and one company that offers programming courses to individuals and companies with around 100 employees. All the participating companies stated they use an agile approach in their software development, but all stated they have their own blend of agile practices. The agile practices ranged from a Water-Scrum-Fall to purer Scrum or Kanban to more XP-like implementations. We asked our company contacts to let us meet teams that they assessed were of different maturity in their collaboration.

We gave all the participating teams a 1.5-hour group development training using the Integrated Model of Group Development with a discussion on the applicability to their own team, i.e.\ where they though their own team was in its development and what they need to develop further. After at around a month we went back to the organizations and asked the participants the following two questions and one statement: 1) \emph{How (if at all) did the workshop on group development psychology influence the teamwork?}, 2) \emph{Did the team mention the group development training at any time during work after the workshop?}, and 3) \emph{Please state the main content of what was being said.} We received 44 individual responses with open feedback in paper form from the 12 teams on the first question, 42 on the second, and 35 on the final statement asking for details. These numbers correspond to the response rates 52\%, 49\%, and 41\%, which is within the range of around 70\% of academic survey studies \citep{baruch2008survey}. 

We then read all the statements obtained and sorted them based on the three categories \emph{positive}, \emph{neutral} or \emph{negative} for the first questions. We then read all the statements and summarized them into different themes and four reoccurring positive themes emerged (see below). The second question mostly triggered \emph{Yes} or \emph{No} answers, but the content of what had been discussed in the teams included many organizational and team aspects of the respondents workplace. We selected quotes that represented as different responses as possible and excluded similar quotes to the ones already selected. Finally, we looked at all the summarized results and assessed what the perceived effects were overall.






\section{Results and Analysis}
\subsection{How (if at all) did the workshop on group development psychology influence the teamwork?}
From the first question 30 out of 44 of the statement in the open feedback were positive, 10 were neutral and four were negative. The results show that the agile teams obtained a \emph{higher awareness of team problems}. Two examples of such feedback are ``We started to think on which level our team is and why, and the reasons why we have problems and how to improve.'' and ``It was a good experience because it affected the way we relate to each other. It got us to look within the team, its problems, positive, and negative points.'' Starting to think about the team as a unit and what it must go through seems to have influenced teams to reflect on why they have some problems and what they might stem from. The workshop also seems to have triggered a reflection on why and how team members relate to each other. 

The second category that surfaced was on \emph{how to deal with team conflict}. Two examples being ``I guess that we are now more open to discussions and people feel more free to show their ideas and not to agree with everything.'' and ``We are now dealing with the conflicts in the team.'' So not only did the workshop increase the awareness and get the teams to deal with their conflicts, it also seems to have helped team members to understand the conflict is a natural, and necessary, part of building a team. If team-related conflict is seen as a part of the process of getting the team together, team members seem to have started thinking that constructive disagreement actually is good, and they dare to express it now. 

The third category was an increased \emph{awareness of what group processes are} from a psychological perspective and how they affect us all in teams, i.e.\ what happens to groups as they progress over time. A couple of examples of how the results show the usefulness of such awareness are: ``We started rethinking how we lead in relation to the interactions we have with developers,'' ``The workshop showed us things we already knew, but didn't know how to address. With higher knowledge on how a group develops, we could improve our weaknesses in the team and also the group work in general,'' ``I changed my view of teamwork,'' ``It helped us see that there's a scientific approach to how this group\slash team works together. It was crazy to see ourselves perfectly described by the group development model.'' and ``We have not been able to discuss it enough, but in this short time, we had relevant discussions about our routines and how we can improve them. We also put some ideas to practice, I think the team is getting stronger. I wish we had more time to discuss this… certainly it would bring great results in the future.''

As can be seen from these quotes, leaders of some teams started reflecting on how they manage developers in relation to team dynamics. In some cases, the teams already knew about their problems, but the workshop gave them tools to actually address them, and one common denominator was that many team members were given a new way to look at teamwork that is also based on scientific research. Teams that have high levels of conflict are also given the possibility to look at their own issues from an outside perspective. Many teams that have not reflected on teamwork get surprised of how well their team fits into the model and it really helps them to feel that their specific team is not strange or in relation to what many teams must go through. Almost all teams were eager to learn more and saw great value in working on the group development in order to improve their teams.

The final category was \emph{better communication}. A couple of examples are ``I think it has helped to bring the group together, a new member has arrived since the workshop and we have tried to make the person comfortable to express opinions,'' ``The workshop helped the team to bond better and further improve the communication and interaction between the members,'' ``The union of the team was improved and now we have team t-shirts :)'' and ``It made us aware of our accomplishments and difficulties. It also helped us to discuss new solutions to one of our problems (communication with customer\slash how to better understand the requirements).'' 

One clear benefit was that the teams learned how to integrate new members better and know what a change in members does to the group dynamics. The importance of building a team spirit seems to also have been a perceived effect and some teams made effort to increase their belonging, like for example creating team t-shirts. Some teams also state that the workshop helped them to deal with external issues and find solutions to them.

The neutral feedback was about no changed to the teamwork in line with ``Hmm… barely. It was very interesting, though,'' ``The communication between the team members has improved, but I'm not sure how it's related to the workshop.'' and ``For me it was good. I don't know how the team was influenced though,'' which indicates that the workshop was in fact interesting but the teams did not reflect more on it. This could be due to the fact that we did not have any followup with the teams. The negative comments were in line with ``It didn't influence us'' or  ``We now know where we are, but it didn't influence us.''



\subsection{Did the team mention the group development at any time during work after the workshop?}
When being asked if the team had mentioned the workshop 31 said \emph{Yes} and 10 said \emph{No}. One person did not respond with a yes or no but instead wrote: ``I can't say that for sure because I haven't been active in the team lately.''

Detailed comments on what had been discussed in the teams after the workshop is shown in the text box. As can be seen the content of what the teams had discussed touched upon a range of topics. Some teams had discussed group development aspects almost every day, which indicated that they find it very useful for improving the team. They generally seem to have realized that all the team members must work together for the team to really excel. Some participants with more managerial tasks within the companies also started reflecting on what group development theory might imply for how they optimize the entire organization. 

They seem to also have realized that building a mature team takes both time and effort and individual group members cannot be seen as just a technical resource but instead as team members with different social identities depending on which team they happen to be in. The workshop also highlighted the importance of clear strategic goals so that all employees know the purpose of the work being conducted. Some teams wrote that they now try to seek clarification of company goals.

\begin{framed}
\scriptsize \textbf{Text box: Examples of detailed comments on what had been discussed.} 
\begin{enumerate}
\item ``Yes, many times! Maybe not everyday, but almost everyday. We even mentioned `stage 4' when we're kidding with each other, e.g.\ `Oh, if you don't eat that, we'll never get to stage 4! Everyone ate it, don't mess up the group, please :)''
\item ``For a long time we kept mentioning and looking forward to get to stage 4, learn and be able to be integrated and develop personal relationships as well as professional ones.''
\item ``Yes, we wish to reach stage 4 with our team, and we know that in order to reach that, all members much work together.''
\item ``The team was eager to use everything from the workshop in relation to what they face as a team.''
\item ``How we can improve other groups and how can we improve our groups in a way so the whole company grows.''
\item ``The significance of communication.''
\item ``How the entire company structure influences the development of the team and everyone is responsible\slash takes part in the success\slash failure while developing a product.''
\item ``We talked about the stages and about our results. We discussed the question\slash practice of separating people from mature teams and discussed some processes of the company.''
\item ``Achieving maturity in a team is not an easy or fast thing to do. We know that and when we feel integrated, accepted within a friendly group, it becomes easier to focus on professional issues.''
\item ``We have discussed about the current group development stage, and how recent members changed (one has left, another has joined) should be dealt with.''
\item ``We said that we wanted to improve the relationship of team members and work more consistently with increased quality. We wanted to know the company goals in more detail.''
\end{enumerate}
\end{framed}

\section{Discussion}
Even though we met the teams for a relatively short time (1.5 hours) we seem to have triggered a reflection around the team dynamics that the team members did not have previously. We therefore have a positive answer you our research question and a first indication of that working on group development is helpful also in agile teams. This can then be seen as a first argument for causality (even though only perceived) between the correlations between group maturity and team agility in software development, as shown in \citet{grenjss2}. The participants perceive these discussions to be due to our workshop for the most part, which means we see great potential in supporting teams with these types of reflection over time. Even the neutral and negative comment were not in relation to how interesting the content of the workshop was, but rather the perceived effects of the workshop on their daily work. The fact that a majority saw such benefits both in relation to their teamwork and what they define as important support from the rest of their organizations (i.e.\ reflection on the company as a whole) show that they perceived an effect that expended the boundaries of the team, which we did not expect. All-in-all, we believe we can state that there were positive effects of the group development psychology training on the agile software development teams that participated in this study, and if we by only a few percent managed to increase the \emph{awareness of team problems}, 
\emph{how to deal with team conflict}, \emph{awareness of what group processes are}, and improve the teams' \emph{communication}, we think these kinds of reflection should be a part of any agile framework.

\subsection{Threats to Validity}
This is a small study conducted on a convenience sample of organizations in Brazil and further studies are needed in order to generalize to a lager population. One of the greatest threats to our study is the lower response rates, and we do not know of any systematic reasons why many team members chose to not respond. Even if they are within 70\% of academic survey studies \citep{baruch2008survey}, response rates around 50\% leaves room for a critical mass of participants actually being of a different view than the replies we obtained. We also do not know if these perceived effects are real effects or if any confounding factors caused the perceived effects of our short workshop. This study should of course be replicated with other measurements of productivity in order to assess the real effects of this training in connection to more support and continuous reflection on these topics. We assess the construct validity as high since we only aimed at investigating the perceived effects of our workshop, i.e.\ what the participants thought in connection to our research question.


\section{Conclusion and Future Work}
In conclusion, the participating teams seem to have a very positive view of group development training and state that they now have a new way of thinking about teamwork, and new tools to deal with team-related problems. We, therefore, see a huge potential in conducting group developmental psychology training with agile software development teams. Helping software development organizations thinking and relating to teams instead of individuals could potentially improve both the effectiveness and well-being of employees, since they get help with dealing with, for example, relationships and conflict. For future work we particularly suggest larger studies in relation to more objective measurements of productivity, and to see if different types of team differ in how much effect the training has. It would also be interesting to see if team size or maturity levels affect how successful the group development training was.


\bibliographystyle{model5-names}
\bibliography{references}
\newpage

\end{document}